\documentclass[12pt]{article}
 \usepackage{graphicx}
\usepackage{epstopdf}
\usepackage{setspace}
\usepackage{amssymb,amsmath,array,tabularx,verbatim}
\usepackage{subfigure}

\usepackage[latin1]{inputenc}
\usepackage[T1]{fontenc}

\newcommand{\bea}{\begin{eqnarray*}}
\newcommand{\eea}{\end{eqnarray*}}

\usepackage{bm}
\usepackage[authoryear]{natbib}
 \usepackage{subfigure}

\newtheorem{Theorem}{\bf Theorem}[section]
\newtheorem{Corollary}{\bf Corollary}[section]
\newtheorem{Remark}{\bf Remark}[section]

\newtheorem{Example}{Example}[section]
\newtheorem{Lemma}{Lemma}[section]

\newcommand{\eff}{\mathrm{eff}}

\newcommand{\be}{\begin{eqnarray}}
\newcommand{\ee}{\end{eqnarray}}

\begin{document}

\parindent 0cm
\title{Designing dose finding studies with an active control for exponential families}

\author{
{\small Holger Dette, Katrin Kettelhake } \\
{\small Ruhr-Universit\"at Bochum } \\
{\small Fakult\"at f\"ur Mathematik } \\
{\small 44780 Bochum, Germany } \\
{\small e-mail: holger.dette@ruhr-uni-bochum.de }\\
\and
{\small Frank Bretz } \\
{\small Statistical Methodology } \\
{\small Novartis Pharma AG } \\
{\small  4002 Basel, Switzerland } \\
{\small e-mail: frank.bretz@novartis.com }\\
}

\maketitle

\begin{abstract}

In a recent paper \cite{detkisbenbre2014} introduced optimal design problems for dose finding studies with an active control. These authors concentrated on regression models with normal distributed errors (with known variance)  and the problem of determining optimal designs for estimating the smallest dose, which achieves the same treatment effect as the active control.
This paper discusses the problem of designing active-controlled dose finding studies from a broader perspective. In particular, we consider a general class of optimality criteria and models arising
from an exponential family, which are frequently used analyzing count data.
 We investigate under which circumstances optimal designs for dose finding studies including a placebo can be used to obtain optimal designs for studies with an active control. Optimal designs are constructed for several situations and the
differences arising from different distributional assumptions are investigated in detail.
In particular, our results are applicable for constructing optimal experimental designs to analyze active-controlled dose finding studies with discrete data,
and we illustrate the efficiency of the new optimal designs with two recent examples from our consulting projects.

\end{abstract}



\section{Introduction}\label{sec1}
\def\theequation{1.\arabic{equation}}
\setcounter{equation}{0}

Dose finding studies are an important tool to investigate the effect of a compound on a response of interest and have numerous applications in various fields such as  medicine, biology or toxicology. They are of particular importance in pharmaceutical drug development because marketed doses have to be safe and provide clinically relevant efficacy  [see \cite{ruberg1995,ting2006}].
Most  of the literature  on statistical methodology for analyzing
 dose response studies include placebo as a control group [see \cite{pinbrebra2006,bretzetal2008}, among others]. Numerous authors have worked on the problem of determining optimal designs for dose response experiments with a placebo group because the application of efficient designs can substantially increase the accuracy of statistical analysis [see \cite{zhuwong2000,fedleo2001,kresmyfung2002,wufedpro2005,drahsupad2007,milguidet2007,borbredet2011}, among many others].\\
However, dose response studies including a marketed drug as an active control are becoming more popular, especially in preparation for an
active-controlled confirmatory non-inferiority trial where the use of placebo may be unethical.
Thus, considerable interest on active-controlled studies
has emerged, as documented through the release of several related guidelines by regulatory agencies
[see  \cite{iche4}, \cite{emea2006, emea2011}, \cite{emea2005}].
Recently \cite{helbenfri2014} investigated the finite sample properties of maximum likelihood estimates of the target dose in an active-controlled study, which achieves the same treatment effect as the active control and \cite{helbenzinknefri2014} studied nonparametric estimates for this quantity.

Despite of these important applications, to our best knowledge, optimal design problems for active-controlled dose finding studies
 have only  been considered in one paper so far [\cite{detkisbenbre2014}].
These authors  investigated   optimal   designs
for estimating the target dose under the assumption of
a normal distribution with known variances. In particular, they demonstrated the superiority of the optimal designs compared to standard designs
used in pharmaceutical practice. However, this
work  is restricted to normal distributed responses with known variances and a special $c$-optimality criterion and obviously the designs derived in their paper are not
necessarily useful for other applications. Therefore the goal of the present paper is to investigate optimal design problems for dose finding studies with an active control from a more general perspective. A first objective is to consider a general class of optimality criteria.
Second, as  it will be pointed out in the following  paragraph, in many dose finding trials with an active control the assumption of normal distributed responses is hard to justify, and we consider
exponential families for modeling the distribution of the responses of the new drug and the active control. This allows in particular to design experiments for controlled studies with discrete data as they have appeared
in  the  consulting projects described in the next paragraph. Third, even if the assumption of a normal distribution is justifiable, we will demonstrate that the estimation of the variances has a nontrivial effect on the optimal designs for an active-controlled study.

The research in the present paper is motivated by two clinical trial examples where the assumption of normal distributed responses made by \cite{detkisbenbre2014} is hard to justify.
The first example refers to a $24$-week, dose-ranging, Phase II study in patients with gouty
arthritis to determine the target dose of  a compound in preventing signs and symptoms of flares in chronic gout patients starting allopurinol
therapy.
The study population consists of male and female patients (age  $18-80$ years) diagnosed with
chronic gout as defined by the American College of Rheumatology preliminary criteria (ACR) and willing to either initiate allopurinol therapy
or having just initiated allopurinol therapy within less than one month. Approximately 500 patients are screened in order to randomize 440
patients in approximately $100$ centers worldwide. Patients who meet the entry criteria are randomized  to receive either the active
comparator or a specific dose  of the new compound. The primary endpoint is the number of
flares occurring per subject within 16 weeks of randomization, which are modeled using a negative binomial distribution for all treatment arms,
{where the corresponding probability is modeled by a
dose-response relationship between the (single) dose groups of the new compound and by  a constant parameter for the comparator.} \\
The second example  is a Phase IIb, multicenter, randomized, double-blind, active-controlled
dose-finding study in the treatment of acute migraine,
as measured by the percentage of patients reporting pain freedom at two hours post-dose.
Approximately $500$ patients are randomized worldwide.
Patients who meet the entry criteria are randomized to receive
either the active comparator or one dose of the new compound.
Once the dose of the new compound is selected,
Phase III studies are conducted to evaluate further the efficacy and safety of the
new compound in the targeted patient population.

In Section \ref{sec2} we give an introduction to optimal design theory for models with an active control under general distributional assumptions. In particular we present results, which relate  optimal designs for dose finding studies with a placebo group
to optimal designs for models with an active control. This methodology is used in  Section \ref{sec3} to construct $D$-optimal designs for dose finding studies with an active control. In Section \ref{sec4} we consider the optimal design  problem for estimating the smallest dose,
which achieves the same treatment effect as the active control. In both sections we investigate the effect  of the distributional assumption on the resulting optimal designs.
In particular, we show that different distributional assumptions (as for example a normal or Poisson distribution used to model continuous or  discrete data) leads
to substantial changes in the structure of the optimal designs.
The Appendix contains the proofs of our main results.

For the sake of brevity this paper is restricted to locally optimal designs
which require a-priori information about the unknown model parameters
[see \cite{chernoff1953}, \cite{fortorwu1992},  \cite{fangheda2008}]. These designs can
be used  as benchmarks for commonly used designs. Moreover,
locally optimal designs serve as basis for constructing optimal designs with respect to more sophisticated optimality
criteria,  which
are robust against a misspecification of the unknown parameters [see \cite{pronwalt1985}  or \cite{chaver1995}, \cite{dette1997},
\cite{imhof2001} among others]. Following this line of research the methodology introduced in the present paper can be further developed to adress uncertainty in the preliminary information for the unknown parameters.

\medskip

\section{Modeling  active-controlled dose finding studies using exponential families} \label{sec2}
\def\theequation{2.\arabic{equation}}
\setcounter{equation}{0}

Consider a clinical trial, where patients  are treated either with
an active control (a standard treatment administered at a fixed dose level) or with
a new drug  using different dose levels in order to investigate the corresponding
dose response relationship.
Given a total sample size $N$, we thus allocate $n_1$
and $n_2=N-n_1$ patients to the two treatments. In addition, we determine the optimal number of different dose levels
for the new drug, the dose levels themselves and the optimal number of patients allocated to each dose level to obtain
the design of the experiment.

More formally, we assume that $k$ different dose levels, say $d_1, \ldots , d_k$,  are chosen in a dose range, say $\mathcal{D} \subset \mathbb{R}^+_0$,
for the new drug (the optimal number
$k$ and the dose levels will be determined
by the choice of the design) and that at each dose level  $d_i$ the experimenter can investigate  $n_{1i}$  patients ($i=1,\ldots ,k$), where $n_1=  \sum_{i=1}^k n_{1i}$
denotes the number of patients treated with the new drug.
The optimal numbers $n_{1i}$, more precisely the optimal proportions $n_{1i}/n_1$, will be determined by the choice of the design.
The corresponding responses at dose level $d_i$ are modeled as realizations of independent real valued
random variables $Y_{ij}$ ($j=1,\ldots , n_{1i}$, $i=1,\ldots , k$). Similarly, the responses of patients treated with the active control are modeled as realizations
of independent real valued random variables $Z_1, \ldots , Z_{n_2}$, where the two samples corresponding to the new drug and active control   are assumed to be independent.
For the statistical analysis we further assume
that the random variables $Z_j$  and $Y_{ij}$
have distributions from an exponential family, where the distributions of the latter depend  on the corresponding dose levels $d_i$, that
is
\begin{eqnarray}
 \label{exp1} f_1(y|d_i, \theta_1) & := & \frac {\partial P_{\theta_1} ^{Y_{i1}}}{\partial \nu}  (y) = \exp \{ c^T_1(d_i,\theta_1)T_1(y)-b_1(d_i,\theta_1) \} h_1 (y)  ,\\
 \label{exp2} f_2(z|\theta_2) 	& := & \frac {\partial P_{\theta_2}^{Z_1}}{\partial \nu} (z) = \exp \{ c^T_2 (\theta_2)T_2(z)-b_2(\theta_2)\} h_2(z) .
 \end{eqnarray}
 Here $\nu $ denotes a $\sigma$-finite measure on the real line, $\theta_1 \in\Theta_1 \subset \mathbb{R}^{s_1}, \ \theta_2 \in \Theta_2 \subset \mathbb{R}^{s_2}$ are unknown parameters
 and we use common terminology  for exponential families [see for example \cite{brown1986}]. In particular, the functions $c_1: \mathcal{D} \times \Theta_1 \to \mathbb{R}^{\ell_1}, \ b_1: \mathcal{D} \times \Theta_1 \to \mathbb{R} , \ c_2: \Theta_2 \to \mathbb{R}^{\ell_2}$ and $b_2:  \Theta_2 \to \mathbb{R}$ are assumed to be twice continuously differentiable where $\tfrac{\partial c_1}{\partial \theta_1}, \tfrac{\partial c_2}{\partial \theta_2} \neq 0$ and $T_1$ and $T_2$ denote $\ell_1$- and $\ell_2$- dimensional statistics defined on the corresponding sample spaces. Additionally, the functions $h_1$ and $h_2$ are assumed to be positive (and measurable).

Throughout this paper let $\kappa$ be a variable indicating whether a patient receives the new drug $(\kappa=0)$ or
the active control ($\kappa=1$) and denote
\begin{equation} \label{desspace}
\mathcal{X}  = (\mathcal{D}  \times \{ 0 \}) \cup \{ (C,1) \}
\end{equation}
as the design space of the experiment, where $\mathcal{D}$ is the dose
range for the new drug, $C$ the dose level of the active control and the second component of an experimental
condition $(d,\kappa) \in \mathcal{X}$ determines the treatment $(\kappa = 0, 1)$.
Straightforward calculation shows that the Fisher information at the point
$(d,\kappa)\in \mathcal{X}$  is given by the matrix
\begin{equation} \label{finfo}
I( (d,\kappa) ,\theta) =
 \begin{pmatrix}  I\{ \kappa =0\}  I_1(d,\theta_1) & \mathbf{0} \\
\mathbf{0} & I\{ \kappa =1\}  I_2(\theta_2)
 \end{pmatrix},
\end{equation}
where $\mathbf{0 } $ denotes a matrix of appropriate dimension with all entries equal to $0$, $\theta=(\theta_1^T,\theta_2^T)^T \in \Theta_1 \times \Theta_2 \subset \mathbb{R}^{s_1+s_2}$ is the vector of all parameters, $I\{ \kappa =0\}$ is  the indicator function of the event $\{ \kappa =0\}  $ and the matrices $I_1$ and $I_2$ are  the Fisher information matrices of the two models \eqref{exp1} and \eqref{exp2}, that is
 \begin{eqnarray} \label{i2}
 I_1(d,\theta_1) &=& \mathbb{E} \Bigl [ \Bigl ( \frac {\partial}{\partial \theta_1} \log f_1 (Y | d,\theta_1) \Bigl ) \Bigl( \frac{\partial}{\partial \theta_1} \log f_1 (Y_{i1} | d_i,\theta_1) \Bigr)^T \Bigr] \nonumber \\
 I_2(\theta_2) &=& \mathbb{E} \Bigl [ \Bigl ( \frac {\partial}{\partial \theta_2} \log f_2 (Z | \theta_2) \Bigl ) \Bigl( \frac{\partial}{\partial \theta_2} \log f_2 (Z_{1} |  \theta_2) \Bigr)^T \Bigr].
 \end{eqnarray}
 where the random variables $Y$ and $Z$ have densities $f_1(y|d,\theta_1)$ and $f_2(z|\theta_2) $ defined by \eqref{exp1} and \eqref{exp2}, respectively.
{Note that the  Fisher information in \eqref{finfo} is block diagonal because of the independence of the samples, as different
patients are either treated with the new drug or  the active control. The following examples illustrate the general terminology.}

\begin{Example} \label{examp1}{\rm In order to demonstrate the different structures of the Fisher information
arising from different distributions of the exponential family we consider several examples.
\begin{itemize}
\item[(a)] \cite{detkisbenbre2014} investigated normal distributed responses with known variances
$\sigma_1^2$ and $\sigma_2^2$ for the new drug and the active control, respectively.
For the expectation of the response of the new drug at dose level $d$ they assumed a nonlinear regression model,
say $\eta (d, \vartheta)$, where $ \vartheta = (\vartheta_0,\dots,\vartheta_s)$,
while it is assumed to be equal to $\mu $ for the active control.
If  the variances are not known and have to be estimated from the data,
we have $\theta_1=(\vartheta_0,\dots,\vartheta_s,\sigma^2_1), \ \theta_2 =(\mu, \sigma^2_2)$ for the parameters in models \eqref{exp1} and \eqref{exp2}, respectively.
Standard calculations show that the Fisher information at a point $(d, \kappa) \in \mathcal{X}$ is given by
\eqref{finfo}, where
\begin{eqnarray} \label{finfnorm1}
	I_1 (d,\theta_1 )
	&=&  \begin{pmatrix}  \tfrac{1}{\sigma_1^2}(\tfrac{\partial}{\partial \vartheta} \eta(d,\vartheta)) (\tfrac{\partial}{\partial \vartheta} \eta(d,\vartheta))^T
	  & \bold{0 }
		 \\\bold{0 }  & \frac{1}{2\sigma_1^4}
		   \end{pmatrix}, ~
I_2 (\theta_2)  = \begin{pmatrix}
		  \frac{1 }{\sigma_2^2}  & 0  \\
		  0 & \frac{  1 }{2\sigma_2^4}  \end{pmatrix}.  ~~~~~~
\end{eqnarray}

\item[(b)] As motivated by the examples in Section~\ref{sec1}, it might be more reasonable to consider a different distribution than a normal distribution
in \eqref{exp1} and \eqref{exp2} to model discrete data.
Assume, for example, a
negative binomial distribution with parameter $r_1 \in \mathbb{N}$ for the number of  failures and a function
$\pi(d,\theta_1) \in (0,1)$ for the probability of a success  of  the new drug (at dose level $d$) and
parameters $r_2 \in \mathbb{N}, \mu \in (0,1)  $  for the active control. Then we have $\theta_2=\mu$ and the Fisher information matrix is given by \eqref{finfo},
where
\begin{eqnarray} \label{finfnbin1}
	I_1 (d,\theta_1 )
	&=& \frac{ r_1  (\tfrac{\partial}{\partial \theta_1} \pi(d,\theta_1))(\tfrac{\partial}{\partial \theta_1} \pi(d,\theta_1))^T}{\pi^2(d,\theta_1) (1-\pi(d,\theta_1 ))} ~, ~
I_2 (\theta_2 )  =
	   \frac{ r_2}{ \mu^2(1- \mu) }. ~~~~
\end{eqnarray}
Here  the parameters $r_1, r_2 \in \mathbb{N}$ for the number of failures are assumed to be known.

\item[(c)] Alternatively, in the case of a  binary response we may  use a Bernoulli distribution,
where $\pi(d,\theta_1) \in (0,1)$ and $ \mu \in (0,1)$ denote the probability of success for the new drug and the active control,
respectively. In this case we have $\theta_2=\mu$ and the Fisher information matrix is given by \eqref{finfo}, where
 \begin{eqnarray*}
 I_1 (d,\theta_1) &=& \frac { ( \tfrac {\partial}{\partial \theta_1} \pi(d,\theta_1) ) ( \tfrac {\partial}{\partial \theta_1} \pi(d,\theta_1) )^T}{\pi(d,\theta_1)(1-\pi(d,\theta_1))} ~,~
 I_2 (\theta_2) = \frac {1}{\mu(1-\mu)}.
 \end{eqnarray*}

\item[(d)]
 For a Poisson distribution, the parameters for the distribution of the responses corresponding to the new drug and the active control
are given by  a function of the dose level, say $ \lambda (d, \theta_1) >0 $, and a parameter  $\mu > 0 $, respectively.
In this case we have $\theta_2=\mu$ and the  Fisher information matrix is given by \eqref{finfo}, where the two non-vanishing blocks are defined by
\begin{eqnarray} \label{finfpoiss1}
	I_1 (d,\theta_1 )
	&=& \frac{  (\tfrac{\partial}{\partial \theta_1}
	\lambda(d,\theta_1)(\tfrac{\partial}{\partial \theta_1} \lambda(d,\theta_1))^T }{\lambda(d,\theta)}  ~,~~
I_2 (\theta_2 )  =
		  \frac {1 }{\mu} .   \label{finfpoiss2}
\end{eqnarray}

\end{itemize}
}
\end{Example}

 Throughout this paper we consider approximate designs in the sense
of \cite{kiefer1974}, which are  defined as  probability measures with finite support on the design space $\mathcal{X}$ in \eqref{desspace}. Therefore,
an experimental design is given by
\begin{equation}\label{desall}
\xi=\begin{pmatrix} (d_1,0)& \dots &(d_k,0) &(C,1)\\
w_1&\dots&w_k&w_{k+1}
\end{pmatrix},
\end{equation}
where $w_1,\dots,w_{k+1}$ are positive weights, such that $\sum^{k+1}_{i=1}w_i=1$.
Here, $w_i$ denotes the relative proportion of patients treated at dose level $d_i \ (i=1, \dots, k)$ or the active control $(i=k+1)$.
  If $N$ observations can be taken, a rounding procedure
is applied
to obtain  integers $n_{1i} $ ($i=1,\ldots,k)$  and $n_2$ from the not necessarily integer valued quantities
$w_iN$ ($i=1,\ldots, k+1$) [see \cite{pukrie1992}]. Thus, the experimenter assigns $n_{11}, \ldots, n_{1k}$ and $n_2$ patients to the dose levels $d_1, \ldots d_k$ of the new drug and the active control, respectively.
In the following discussion we will determine optimal designs, which also optimize the number $k$ of different dose levels. It turns out that for the models considered here the optimal designs usually allocate observations at less than $5$ dose levels. Note that in  practice the number $k$ of different dose levels is in the range of $4-7$ and rarely  larger than $10$.

   The information matrix of an approximate design $\xi$ of the form \eqref{desall} is
defined by the $(s_1 + s_2) \times (s_1+s_2 )$ matrix
\begin{align} \label{inf}
M(\xi,\theta)&=\int_{\mathcal{X}}  I((d,\kappa) , \theta) d\xi(d,\kappa) =  \begin{pmatrix} (1-\omega_{k+1})   M_1(\tilde \xi, \theta_1) & \bold{0} \\ \bold{0} & \omega_{k+1}
I_2 (\theta_2). \end{pmatrix}
                                    \end{align}
Here, the  $s_1\times s_1$ matrix $ M_1(\tilde \xi, \theta_1)$  and the $s_2 \times s_2$ matrix $ I_2(\theta_2)$ are
 given by
\begin{eqnarray}\label{tildea}
M_1 (\tilde \xi,\theta_1) &=& \int_{\mathcal{D}}  I_1(d,\theta_1) d \tilde \xi(d) ,
\end{eqnarray}
and \eqref{i2},  respectively, and
\begin{equation}\label{tilde}
\tilde{\xi}=\begin{pmatrix} d_1&\dots&d_k\\\tilde w_1 &\dots&\tilde w_k \end{pmatrix}
\end{equation}
denotes the design (on the design space $\mathcal{D}$) for the new drug, which is induced by the design $\xi$ in \eqref{desall} defining the weights $ \tilde w_i =\frac{w_i}{1-w_{k+1}}$, $i=1, \ldots, k$.

 If observations are taken according to an approximate design it can be shown (assuming standard regularity conditions) that the maximum likelihood estimators $\hat \theta_1, \hat\theta_2$ in models \eqref{exp1} and \eqref{exp2} are asymptotically normal distributed, that is
 $$
 \sqrt{N} \big( (\hat \theta_1^T, \hat \theta_2^T)^T - (\theta^T_1, \theta^T_2) ^T \big) \stackrel{ \mathcal{D}}{\longrightarrow} \mathcal{N} (\bold{0}, M^{-1}(\xi,\theta))
 $$
 as $N {\longrightarrow}\infty$,
where the symbol $\stackrel{ \mathcal{D}}{\longrightarrow}$ denotes convergence in distribution. \cite{debrpepi2008} considered   dose finding studies including a placebo group and
 showed by means of a simulation study that the approximation of the variance of $\hat \theta =  (\hat \theta_1^T, \hat \theta_2^T)^T $ by $\tfrac{1}{N}M^{-1}(\xi,\theta)$ is satisfactory for total sample sizes larger than $25$. As typical clinical dose finding trials have sample sizes in the range of
 $200 - 300$ [see for example \cite{phrma:2007}], it is reasonable to  use this  approximation also for active-controlled studies. Consequently, optimal designs maximize an appropriate functional of the information matrix defined in \eqref{inf}.

 In order to discriminate between competing designs we consider in this paper Kiefer's $\phi_p$-criteria [see \cite{kiefer1974} or \cite{pukelsheim2006}]. To be precise, let $ p \in [-\infty,1)$ and $ K \in \mathbb{R}^{(s_1 + s_2) \times t}$ denote a matrix of full column rank $t$. Then a design $\xi^*$ is called locally $\phi_p$-optimal for estimating the linear combination $K^T \theta$
in a dose response model with an active control, if $K^T \theta$ is estimable by the design $\xi^*$, that is, $K^T \theta \in$ Range$(M(\xi^*,\theta))$, and $\xi^*$ maximizes the functional
 \begin{equation} \label{crit}
 \phi_p(\xi) =  \Bigl(\frac {1}{t} \mbox{tr} (K^T M^- (\xi,\theta)K)^{-p} \Bigr)^{\frac {1}{p}}
 \end{equation}
 among all designs for which $K^T \theta$ is estimable, where tr$(A)$ and $A^-$ denote the trace and a generalized inverse of the matrix $A$, respectively.
 Note that the cases $p=0$ and $p=-\infty $ correspond to the $D$- and $E$-optimality criterion, that is
 $ \phi_0(\xi)  = \det (K^T M^- (\xi,\theta)K )^{-\tfrac{1}{t}}$ and   $ \phi_{-\infty} (\xi)  = \lambda_{\min} ((K^T M^- (\xi,\theta)K) ^{-1})$.
 An application of the general equivalence theorem [see \cite{pukelsheim2006}, chapter 7.19 and 7.21, respectively] to the situation considered in this paper yields immediately the following result.

 \begin{Lemma}\label{lem1} If $p \in (-\infty , 1)$, a
 design $\xi^*$ with $K^T \theta \in$ Range$(M(\xi^*,\theta))$ is locally $\phi_p$-optimal for estimating the linear combination $K^T \theta$ in a dose response model with an active control if and only if there exists a generalized inverse $G$ of the information matrix $M(\xi^*,\theta)$, such that  the inequality
{\small{
\begin{equation}\label{aequ}
{ \rm tr} \bigl( I((d,\kappa),\theta)GK(K^TM^-(\xi^*,\theta)K)^{-p-1} K^TG^T\bigr)  - { \rm tr} (K^TM^-(\xi^*,\theta)K)^{-p} \leq 0
\end{equation}
}}
holds for all $(d,\kappa ) \in \mathcal{X}$.
If $p = -\infty  $, a design $\xi^*$ with $K^T \theta \in$ Range$(M(\xi^*,\theta))$  is
locally $\phi_{-\infty}$-optimal for estimating the linear combination $K^T \theta$
if and only if there exist a generalized inverse $G$ of the information matrix $M(\xi^*,\theta)$ and a nonnegative definite matrix $E \in \mathbb{R}^{t \times t}$ with $\mbox{\rm tr} (E) =1$,
such that  the inequality
{\small{
\begin{equation}\label{aequ1}
{ \rm tr} \bigl( I((d,\kappa),\theta)GK(K^TM^-(\xi^*,\theta)K)^{-1} E (K^TM^-(\xi^*,\theta)K)^{-1} K^TG^T\bigr)  - \lambda_{\min}((K^TM^-(\xi^*,\theta)K)^{-1}) \leq 0
\end{equation}
}}
holds for all $(d,\kappa ) \in \mathcal{X}$.
Moreover, there is equality in \eqref{aequ}  ($p> - \infty$) and  \eqref{aequ1}  ($p=-\infty$) for all support points of the design $\xi^*$.
 \end{Lemma}

In the following discussion we assume that either $p=-1$ or that the matrix $K$ is a block matrix of the form
\begin{equation}\label{kdiag}
K= \left( \begin{array}{cc}
K_{11} & 0 \\ 0 & K_{22}
\end{array}  \right) \in \mathbb{R}^{(s_1 + s_2) \times (t_1 + t_2)}
\end{equation}
with elements $K_{11} \in \mathbb{R}^{s_1 \times t_1} , \ K_{22} \in \mathbb{R}^{s_2 \times t_2}$, $t_1+t_2=t$. Roughly speaking, the choice $p=-1$ or a blockdiagonal structure of the matrix $K$ in \eqref{kdiag}
leads to a separation of the parameters from models \eqref{exp1} and \eqref{exp2} in the corresponding optimality criterion. As a consequence
optimal designs for dose finding studies with an active control can be obtained from
optimal designs for dose finding studies including a placebo group, which maximize the criterion
\begin{equation} \label{crittilde}
\tilde \phi_p (\tilde \xi) = \Bigl( \frac {1}{t_1} {\rm tr} (K^T_{11}M^-_1(\tilde \xi,\theta_1)K_{11})^{-p}\Bigr)^{\frac {1}{p}}
\end{equation}
in the class of all designs $\tilde \xi$ for which $K^T_{11}\theta_1$ is estimable, i.e. $K_{11}^T \theta_1 \in \mbox{Range}( M_1(\tilde \xi,\theta_1))$. Throughout this paper these designs are called $\tilde \phi_p$-optimal for estimating the parameter $K_{11}^T\theta_1$ in the dose response model \eqref{exp1}. The proof can be found in the Appendix.

\medskip

\begin{Theorem} \label{thm1}
Assume $p \in [-\infty , 1)$, that the matrix $K$ is given by \eqref{kdiag} and that
\begin{eqnarray} \label{tilde}
\tilde \xi^*_p= \left( \begin{array}{ccc}
d^*_1 & \dots & d^*_k \\
\tilde w^*_1 & \dots & \tilde w^*_k
\end{array} \right)
\end{eqnarray}
is a locally $\tilde \phi_p$-optimal design for estimating $K^T_{11}\theta_1$ in the dose response model \eqref{exp1}. Then the design
\begin{eqnarray*}
\xi^*_p = \left ( \begin{array} {cccc}
(d^*_1,0) & \dots & (d^*_k,0) & (C,1) \\
w^*_1  & \dots & w^*_k & w^*_{k+1}
\end{array}\right )
\end{eqnarray*}
is locally $\phi_p$-optimal for estimating $K^T\theta$ in the dose response model with an active control, where the weights are given by
\begin{equation} \label{weight}
w^*_{k+1} = \frac {1}{1+\rho_p} , \qquad w^*_i = \frac {\rho_p}{1+ \rho_p} \tilde w_i^* \qquad (i=1,\dots,k),
\end{equation}
and
\begin{equation} \label{rho}
\rho_p = \frac {({\rm tr}((K^T_{22}I^-_2(\theta_2)K_{22})^{-p}))^{1/(p-1)}}
{({\rm tr}((K^T_{11}M^-_1(\tilde \xi^*_p, \theta_1)K_{11})^{-p}))^{1/(p-1)}}
\end{equation}
(the case $p=-\infty$ is interpreted as the corresponding limit).
\end{Theorem}

\medskip

In the case $p=-1$ a more general statement is available without the restriction to block matrices of the form \eqref{kdiag}. The proof is obtained by similar arguments as presented in the proof of Theorem \ref{thm1} and therefore omitted.

\begin{Theorem} \label{thm2}
Assume that $K^T=(K^T_{11}, K^T_{22}) \in \mathbb{R}^{t \times (s_1+s_2)}$ with $K^T_{11} \in \mathbb{R}^{t \times s_1}, K^T_{22} \in \mathbb{R}^{t \times s_2}$ and let $\tilde \xi_{-1}^*$ denote the $\tilde \phi_{-1}$-optimal design for estimating the parameter $K_{11}^T \theta_1$ in the dose response model \eqref{exp1}. Then the design $\xi^*_{-1}$ defined in Theorem \ref{thm1} is locally $\phi_{-1}$-optimal for estimating $K^T \theta$ in the dose response model with an active control.
\end{Theorem}

The final result of this section considers the special case $p=0$. The result is a direct consequence of Theorem \ref{thm1} considering the limit $p \to 0$ and observing that the quantity $\rho_p$ defined in $\eqref{rho}$ satisfies
$
	\lim_{p \to 0} \rho_p = \tfrac{t_1}{t_2}.
$

\begin{Corollary} \label{cor1}
	Assume that the matrix $K$ is given by \eqref{kdiag} and let $\tilde \xi_0^*$ denote the locally $D$-optimal design of the form \eqref{tilde} for estimating the parameter $K^T_{11} \theta_1$ in the dose response model \eqref{exp1}, which maximizes $\det((K_{11}^T  M_1^-(\tilde \xi,\theta_1)K_{11})^{-1})$ in the class of all designs for which $K_{11}^T \theta_1$ is estimable. Then the design
	\begin{equation}
		\xi_\theta^* = \left ( \begin{array}{cccc}
(d^*_1,0) & \dots & (d^*_k,0) & (C,1)
\\
\tfrac{t_1}{t_1+t_2} \tilde w^*_1 & \dots & \tfrac{t_1}{t_1+t_2} \tilde w_k^* & \tfrac{t_2}{t_1+t_2} \end{array} \right )
	\end{equation}
is locally $D$-optimal for estimating the parameter $K^T \theta$ in the dose response model with an active control.
\end{Corollary}

\begin{Remark} {\rm  The assumption of a  block matrix  $K$  in Theorem  \ref{thm1} and Corollary \ref{cor1} can not be omitted.
Consider for example the case of a binomial distribution and  a Michaelis-Menten model   $\pi(d,\theta_1)=\tfrac{\vartheta_1 d}{\vartheta_2 +d}$
for the dose response relationship of the new drug. Assume that one is interested in two functionals of the model parameters:
(i) The  estimation of the distance between the effect of the active control, say  $\theta_2$, and the effect of the new drug at a special dose level $d_0$, i.e.  $\pi (d_0,\theta_1)$,
and (ii) the difference between $\theta_2$ and the maximum effect of the new drug, i.e. $\vartheta_1$.
 In this case the matrix $K$ is given by
$$
K=\begin{pmatrix} -\tfrac{d_0}{\vartheta_2+d_0} & \tfrac{\vartheta_1 d_0}{(\vartheta_2+d_0)^2}& 1 \\ -1 & 0 & 1 \end{pmatrix}^T.
$$
Consider exemplarily  the choice  $\mathcal{D}=[0,50]$, $\vartheta_1=0.5$, $\vartheta_2=2$, $\theta_2=0.4$ and $d_0=5$.
The  locally $D$-optimal design  for estimating
$K^T\theta$ in the dose response model with an active control
allocates $36\%$ and $32\%$ of the patients to the dose levels $0.93$ and $50$ of the new drug and $32\%$ of the patients to the active control, respectively. The corresponding function
 \eqref{aequ}
of  the equivalence theorem is shown in the left panel of Figure \ref{equAC}  for the case $\kappa=0$.
In the case $\kappa=1$ this function reduces to the constant $0$ for all $d \in \mathcal{D}$.
\begin{figure}
\tiny
\centering
	\subfigure{\includegraphics[scale=0.4]{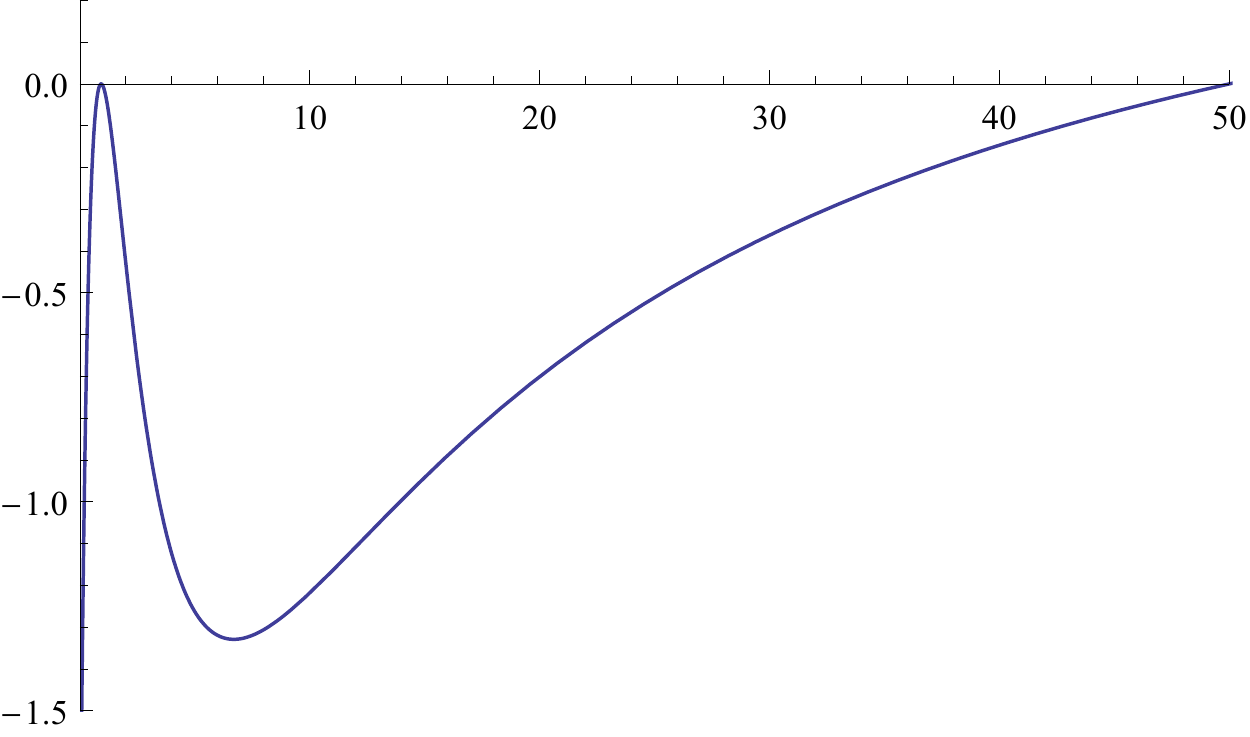}}
~~
	\subfigure{\includegraphics[scale=0.4]{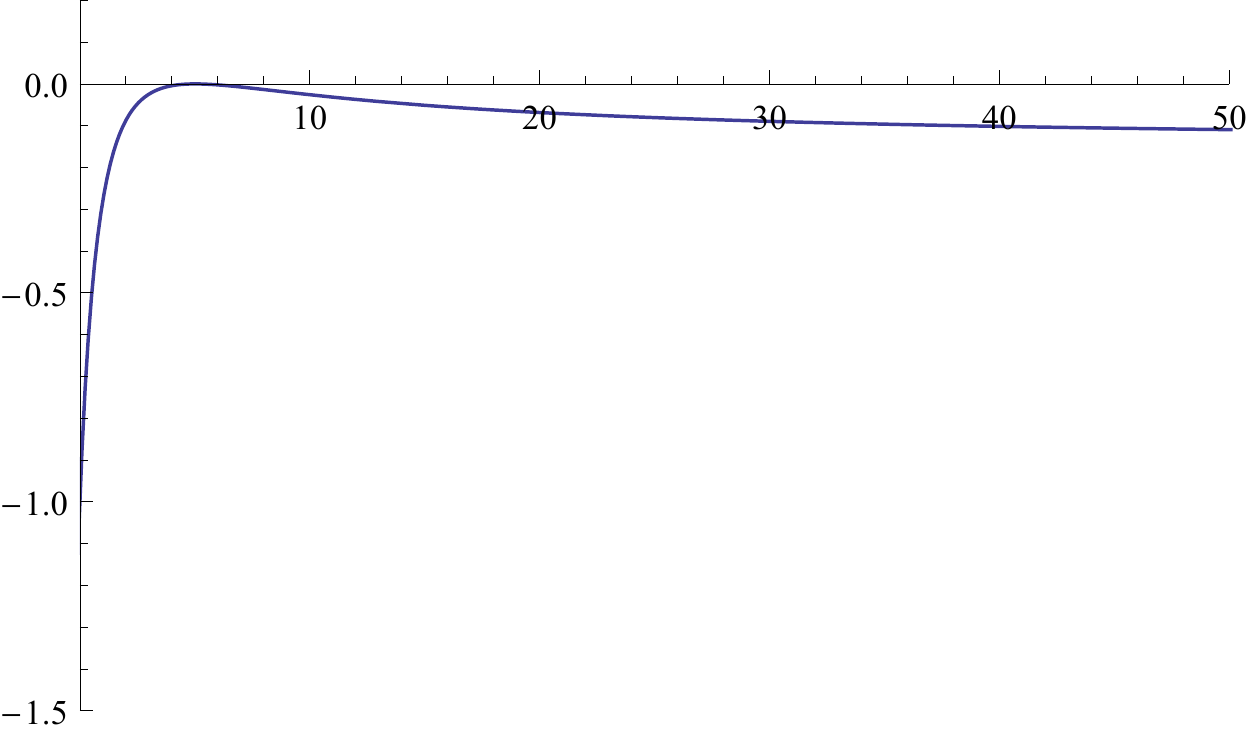}}
~
	\subfigure{\includegraphics[scale=0.4]{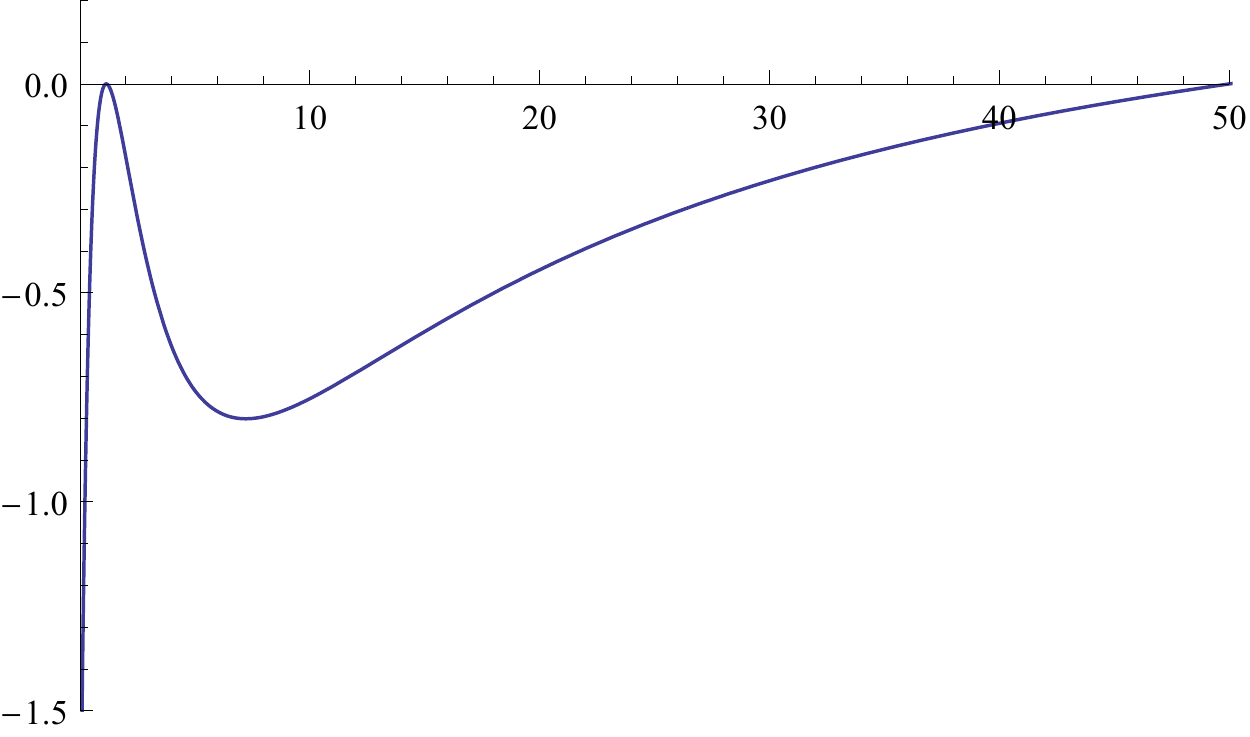}}
\caption{\it The inequality \eqref{aequ} of the equivalence theorem. Left panel: $D$-optimal design for estimating $K^T\theta$ in the dose response model with an active control. Middle
panel and right panel: optimal design for prediction and $D$-optimal design in the model \eqref{exp1}.}
\label{equAC}
\end{figure}
In the situation where no active control is available one could look at the
problem designing the experiment for a most efficient {estimation} of
 $\pi (d_0,\theta_1)$. This corresponds to the
 matrix  $K_{11}=(\tfrac{d_0}{\vartheta_2+d_0} , -\tfrac{\vartheta_1 d_0}{(\vartheta_2+d_0)^2})^T$ and
the  locally optimal design  is a one point design which treats $100\%$ of the patients with the dose level $d_0=5$.
On the other hand the locally $D$-optimal design for estimating $\theta_1$ allocates
$50\%$ of the patients to each of the dose levels  $1.15$ and  $50$.
The corresponding inequalities of the equivalence theorem are  shown in the middle and right panel of Figure \ref{equAC}.
Obviously the locally $D$-optimal design  for estimating
$K^T\theta$ in the dose response model with an active control can not be derived from these designs and an assumption of the type  \eqref{kdiag} is in fact necessary to obtain Theorem \ref{thm1}.
}
\end{Remark}

\section{$D$-optimal designs for the Michaelis-Menten and EMAX model}
\label{sec3}
\def\theequation{3.\arabic{equation}}
\setcounter{equation}{0}

In this section we determine some $D$-optimal designs for dose finding studies with an active control under different distributional assumptions. We assume that the dependence on the dose of the new drug is either described by the Michaelis-Menten model
$\frac {\vartheta_1d}{\vartheta_2+d},$
or the EMAX model
$\vartheta_0 + \frac {\vartheta_1d}{\vartheta_2+d},$
where the dose range is given by the interval $[L,R] \subset \mathbb{R}^+_0$. These models are widely used
when investigating the dose response relationship
of a new compound, such as a medicinal drug, a fertilizer, or an environmental toxin.
Note that in the case where the function describes a probability, one requires some restrictions on the parameters. For example, if $\pi(d,\theta_1)= \frac {\vartheta_1d}{\vartheta_2+d}$ is the probability of a success for the negative binomial distribution in Example \ref{examp1}(b), we implicitly assume $\frac {\vartheta_1R}{\vartheta_2+R}<1$ in the following discussion. In other models similar assumptions have to be made and we do not mention these restrictions explicitly for the sake of brevity. In the following, $x \vee y$ denotes the maximum of $x, y \in \mathbb{R}$.

\medskip

\begin{Theorem} (Michaelis-Menten model) \label{thm31a}
\begin{itemize}
\item[(a)]
If the distributions of the responses corresponding to the new drug and active control are normal with parameters $(\frac {\vartheta_1d}{\vartheta_2+d}, \sigma^2_1)$ and $(\mu, \sigma^2_2)$, respectively, then the locally $D$-optimal design for the dose response model with an active control allocates $30\%$ of the patients to each of the dose levels $L \vee \frac {\vartheta_2R}{2\vartheta_2+R}$ and $R$ of the new drug and $40\%$ to the active control.

\item[(b)]
In the case of  negative binomial distributions with probabilities $\pi(d,\theta)=\frac {\vartheta_1d}{ \vartheta_2+d}$ and $\mu$ the locally $D$-optimal design for the dose response model with an active control allocates $33.\overline{3}\%$ of the patients to each of the dose levels $L$ and $R$ of the new drug and $33.\overline{3}\%$ to the active control.

\item[(c)]
In the case of binomial distributions with probabilities $\pi(d,\theta)=\frac {\vartheta_1d}{ \vartheta_2+d}$ and $\mu$ the locally $D$-optimal design for the dose response model with an active control allocates $33.\overline{3}\%$ of the patients to each of the dose levels $L \vee \tfrac{\vartheta_2 R + 3 \vartheta_2^2-\vartheta_2 \sqrt{9 R^2 - 8R^2\vartheta_1+18R\vartheta_2-8R\vartheta_1\vartheta_2+9\vartheta_2^2}}{4\vartheta_1\vartheta_2-4R+4R\vartheta_1-6\vartheta_2}$ and $R$ of the new drug and $33.\overline{3}\%$ to the active control.

\item[(d)]
If Poisson distributions with parameters $\lambda(d,\theta_1)= \frac {\vartheta_1d}{\vartheta_2+d}$ and $\mu$ are used in \eqref{exp1} and \eqref{exp2}, the locally $D$-optimal design for the dose response model with an active control allocates $33.\overline{3}\%$ of the patients to each of the dose levels $L \vee \frac {\vartheta_2R}{3\vartheta_2+2R}$ and $R$ of the new drug and $33.\overline{3}\%$ to the active control.

\end{itemize}
\end{Theorem}

\medskip

The proof of Theorem \ref{thm31a} is a direct consequence of Corollary \ref{cor1}, if the locally $D$-optimal designs for model \eqref{exp1} are known. For example, in the case of a normal distribution it follows from \cite{rasch1990} that the $D$-optimal design for the Michaelis-Menten model has equal masses at the points $L \vee \frac {\vartheta_2R}{2\vartheta_2+R}$ and $R$ and Corollary \ref{cor1} yields part $(a)$ of Theorem \ref{thm31a}. In the other cases the $D$-optimal designs for model \eqref{exp1} are not known and the proof can be found in the Appendix. \\
{ It is also worthwhile to note that the differences of the $D$-optimal designs
derived under different distributional assumptions can be substantial. For example if the design space is $[0,R]$ with  a  large
  right boundary $R$, the non trivial dose level  for the new drug is approximately $\vartheta_2$
and $\vartheta_2/2$ under the assumption of a normal and Poisson distribution, respectively.} We will now give the corresponding results for the EMAX model.
The proof follows by similar arguments as given in the proof of Theorem \ref{thm31a} and is therefore omitted.

\begin{Theorem} (EMAX-model) \label{thm32a}
\begin{itemize}
\item[(a)]
If the distribution of responses corresponding to the new drug and active control are normal distributions with parameters $(\vartheta_0 + \frac {\vartheta_1d}{\vartheta_2+d}, \sigma^2_1)$ and $(\mu, \sigma^2_2)$, respectively, then the locally $D$-optimal design for the dose response model with an active control allocates $22.\overline{2}\%$ of the patients to each of the dose levels $L$, $d^*=\tfrac{R(L+\vartheta_2) + L(R+\vartheta_2)}{(L+\vartheta_2)+(R+\vartheta_2)}$ and $R$ of the new drug and $33.\overline{3}\%$ to the active control.
\item[(b)]
In the case of negative binomial distributions with probabilities $\pi(d,\theta)=\vartheta_0 + \frac {\vartheta_1d}{\vartheta_2+d}$ and $\mu$, the locally $D$-optimal design for the dose response model with an active control allocates $25\%$ of the patients to each of the dose levels $L$, $d^*$ and $R$ of the new drug
and $25\%$ to the active control,  where  $d^*$ is the solution of the equation $$
\tfrac{2}{d-L}+\tfrac{2}{d-R}-\tfrac{\vartheta_0+\vartheta_1-1}{d (\vartheta_0+\vartheta_1-1)+(\vartheta_0-1) \vartheta_2}-\tfrac{2 (\vartheta_0+\vartheta_1)}{\vartheta_0 (\vartheta_2+d)+\vartheta_1 d}-\tfrac{1}{\vartheta_2+d}=0$$
\item[(c)]
In the case of binomial distributions with probabilities $\pi(d,\theta)=\vartheta_0 + \frac {\vartheta_1d}{\vartheta_2+d}$ and $\mu$, the locally $D$-optimal design is of the
same form as described  in part (b), where $d^*$ is the solution of the equation
$$\tfrac{2}{d-L}+\tfrac{2}{d-R}-\tfrac{\vartheta_0+\vartheta_1-1}{d (\vartheta_0+\vartheta_1-1)+(\vartheta_0-1) \vartheta_2}-\tfrac{\vartheta_0+\vartheta_1}{\vartheta_0 (\vartheta_2+d)+\vartheta_1 d}-\tfrac{2}{\vartheta_2+d}=0.$$
\item[(d)]
If Poisson distributions with parameters $\lambda(d,\theta_1)= \vartheta_0 + \frac {\vartheta_1 d}{\vartheta_2+d}$ and $\mu$ are used in \eqref{exp1} and \eqref{exp2}, respectively, then the locally $D$-optimal is
of the same form as described  in part (b), where
$$
d^*=\vartheta_2 \tfrac {4m(L)m(R)-\vartheta_1 (Lm(R)+Rm(L))-\vartheta_0 \sqrt{\kappa}}{-4m(L)m(R)- \vartheta_1 \vartheta_2 (m(R)+m(L))+(\vartheta_1 + \vartheta_0) \sqrt{\kappa}}$$
and $\kappa = ( (\vartheta_2+L)m(R)+(\vartheta_2+R)m(L))^2 + 12 (\vartheta_2+L)(\vartheta_2+R)m(R)m(L)$, $m(d)=\vartheta_0\vartheta_2+\vartheta_1d+\vartheta_0d$.

\end{itemize}
\end{Theorem}

\begin{Example}{\rm
	Under the assumption of a normal distribution, \cite{detkisbenbre2014} determined a $D$-optimal design for the EMAX dose response model ignoring the effect caused by estimating the variance. It follows from Theorem 4 in \cite{detkisbenbre2014} that the locally $D$-optimal design allocates $26.\overline{6}\%$ of the patients to the dose levels $L, d^*, R$ of the new drug and $20\%$ to the active control, respectively, where $d^*$ is defined in 3.2(a). Theorem \ref{thm32a} above shows that the design which accounts for the problem of estimating the variances uses the same
	dose levels but allocates $ 13.3\%$ more patients to the active control. }
\end{Example}

\begin{Example} \label{dataex}{\rm
	In this example we discuss D-optimal designs for the two clinical trials considered in Section~\ref{sec1}.
\begin{itemize}
\item[(a)]
We first consider the gouty arthritis example. The primary endpoint is modeled by a negative binomial distribution with parameters $r_1$ and $\pi(d,\theta_1) = \vartheta_0+\tfrac{\vartheta_1 d}{\vartheta_2 + d}$ for the new drug and parameters $r_2$ and $\theta_2$ for the comparator. The dose range is $[0,300]$mg
and we obtained from the clinical team  the following  preliminary information
for the unknown parameters:  $\vartheta_0=0.26$, $\vartheta_1=0.73$, $\vartheta_2=10.5$, $\sigma_1=0.05$ and $\theta_2=0.9206$, $\sigma_2=0.05$. In addition, $r_1=r_2=10$ are  fixed.
The  $D$-optimal design is obtained from Theorem \ref{thm32a}  and  depicted in the  upper part of Table \ref{tabDeffg}. It
allocates $25\%$ of all  patients to the active control and $25\%$ of the patients to the dose levels $0, 8.23, 300$mg of the new drug, respectively.
 The standard design actually used in this study allocates $14.3\%$ of the patients to the dose levels $25,50,100,200,300$mg of the new drug and $28.5\%$ of the patients to the active control.
To compare these designs we also show  in the last column of Table \ref{tabDeffg}
  the D-efficiency
\begin{equation}
\eff_D(\xi,\theta)=\frac{\Phi_0 (\xi,\theta)}{\Phi_0(\xi_{D}^{*},\theta)} \in [0,1],
\end{equation}
where $\xi_{D}^{*}$ is the locally D-optimal design. We observe that in this example an optimal design improves the
standard  design
substantially. We also observe that the differences between the  $D$-optimal designs calculated under a different distributional assumption are rather small.
For this  example, the $D$-optimal design calculated under the assumption of a normal distribution has efficiency $0.98$ in the model based on the negative binomial distribution.

\renewcommand{\arraystretch}{1.2}
\begin{table}[h]
 \footnotesize
\centering{
\begin{tabular}{|l|c|c|}
\hline
  distribution & D-optimal design  & $\eff_D$ \\
	\hline
  normal &  \begin{tabular}{cccc} (0,0) & (9.81,0) & (300,0) & (C,1) \\ \hline  $22.\overline{2}\%$ & $22.\overline{2}\%$ & $22.\overline{2}\%$ & $33.\overline{3}\%$ \end{tabular} & 0.25 \\
	\hline
	negative binomial  & \begin{tabular}{cccc} (0,0) & (8.23,0)  & (300,0) & (C,1) \\ \hline  $25\%$ & $25\%$ & $25\%$ & $25\%$ \end{tabular} & 0.11 \\
\hline
\hline
  normal &  \begin{tabular}{cccc} (0,0) & (10.95,0) & (200,0) & (C,1) \\ \hline  $22.\overline{2}\%$ & $22.\overline{2}\%$ & $22.\overline{2}\%$ & $33.\overline{3}\%$ \end{tabular} & 0.84 \\
	\hline
	 binomial  & \begin{tabular}{cccc} (0,0) & (9.05,0)  & (200,0) & (C,1) \\ \hline  $25\%$ & $25\%$ & $25\%$ & $25\%$ \end{tabular} & 0.86 \\
\hline
\end{tabular}
}
 \caption{\small \it $D$-optimal designs in the two clinical trials discussed in Section~\ref{sec1} under different distributional assumptions. Upper part:  gouty arthritis example;
 lower part: acute migraine example. The last column shows the efficiencies of the designs, which were actually used in the study.}
 \label{tabDeffg}
 \end{table}

\item[(b)]
We now consider the acute migraine example, which measured the percentage of patients reporting pain freedom at two hours post-dose.
We assume a binomial distribution for this example. The probabilities of success are  $\pi(d,\theta_1)= \vartheta_0+\tfrac{\vartheta_1 d}{\vartheta_2 + d}$
   for the new compound (where  the dose level varies in the interval  $[0,200]$mg)  and  $\theta_2$ for the active control.
   The sample sizes are $n_1=517$ and $n_2=100$ and  the preliminary information obtained from the clinical team is given by
    $\vartheta_0=0.098,$ $ \vartheta_1=0.2052,$ $\vartheta_2=12.3,$  $\sigma_1=0.05$ and $\theta_2=0.2505$, $\sigma_2=0.05$. The locally D-optimal designs under a normal and binomial distribution assumption are listed in the lower part of Table \ref{tabDeffg}. The design actually used for this study allocated $21, 5, 7, 10, 10, 11, 10, 10 \%$  of the patients to the dose levels $0,2.5,5,10,20,50,100,200$mg of the new drug and $16\%$ of the patients to the active control, respectively. The second column of Table \ref{tabDeffg} displays its efficiencies relative to the proposed designs and again a substantial  improvement can be observed under both distributional assumptions.
{For this example, the $D$-optimal design calculated under the assumption of a normal distribution has also efficiency $0.98$ in the model based on the binomial distribution.}
\end{itemize}

}
\end{Example}

\section{Optimal designs for estimating the target dose}
\label{sec4}
\def\theequation{4.\arabic{equation}}
\setcounter{equation}{0}

In this section we investigate the problem of constructing locally optimal designs for estimating
 the treatment effect of the active control and the target dose, that is
the smallest dose of the new compound which achieves the same treatment effect as the active control.
For this purpose we  consider a dose range of the form $\mathcal{D}=[L,R]$ and introduce  the notation
\begin{eqnarray}\label{ex1}
\mathbb{E}_{\theta_1} [Y_{ij}|d_i] &=& \eta(d_i,\theta_1) \qquad  \qquad  (j=1,\dots,n_{1i}, \ i=1,\ldots,k) \\
\label{ex2}
\mathbb{E}_{\theta_2}[Z_i] &=& \Delta \qquad  \qquad \qquad \ \ \ (i=1,\ldots,n_2)
\end{eqnarray}
for the expected values of responses corresponding to the new drug (for dose level $d_i$) and the active control, respectively. We assume
(for simplicity) that the function $\eta$ in \eqref{ex1} is strictly increasing in $d\in \mathcal{D}$ and that $d^*(\theta)=\eta^{-1}(\Delta,\theta_1)$, is an element of the dose range $\mathcal{D}=[L,R]$ for the new drug. Note that the expectation $\Delta$ in \eqref{ex2} is a function of the $s_2$-dimensional parameter $\theta_2$, say $\Delta =k(\theta_2)$. Consequently,
a natural estimate of $d^{\ast}$ is given by
$\hat{d}^{\ast} = d^*(\hat  \theta) =\eta^{-1}(\hat{\Delta},\hat{\theta}_1)$, where $\hat{\Delta} = k (\hat \theta_2)$ and
$\hat{\theta}=( \hat{\theta}_1^T, \hat{\theta}_2^T)^T$ denotes the vector of  the maximum likelihood estimates  of
the parameter $\theta_1 $ and $\theta_2 $ in models  \eqref{exp1}  and \eqref{exp2}, respectively.
Standard calculations show  that the variance of this  estimator  is approximately given by
\begin{equation}\label{approx}
	 \mbox {Var}(d^*(\hat \theta)) \approx \tfrac{1}{N} \psi(\xi,\theta),
\end{equation}
where the function $\psi$ is defined by
\begin{equation}\label{critpsi}
	\psi(\xi,\theta) = \tfrac{1}{1-\omega_{k+1}} (\tfrac{\partial}{\partial \theta_1} d^*(\theta))^T  M^-_1(\tilde \xi,\theta_1) (\tfrac{\partial}{\partial \theta_1}d^*(\theta)) + \tfrac{1}{\omega_{k+1}} (\tfrac{\partial}{\partial \theta_2} d^*(\theta))^T {I}_2  ^-(\theta_2) (\tfrac{\partial}{\partial \theta_2} d^*(\theta_2)),
\end{equation}
$\tilde \xi$ denotes the design for the new drug induced by the design $\xi$, see \eqref{tilde},
 and  $M^-_1(\tilde \xi,\theta_1)$   and $I_2^-(\theta_2)$
are  generalized inverses of the information matrices  $ M_1(\tilde \xi, \theta_1)$ and $I_2( \theta_2)$, respectively.

Following
 \cite{detkisbenbre2014},  we call  a design $\xi^*_{AC}$ locally AC-optimal design (for \underline{A}ctive \underline{C}ontrol) if $\tfrac{\partial}{\partial \theta_1}d^*(\theta) \in \mbox{Range}
(M_1(\tilde \xi,\theta_1))$,   $\tfrac{\partial}{\partial \theta_2}d^*(\theta) \in \mbox{Range}
(I_2(\theta_2))$ and if $\xi^*_{AC}$  minimizes the function $\psi(\xi,{\theta})$ among all designs satisfying this
estimability  condition. Note that the criterion  \eqref{critpsi} corresponds to a $\phi_{-1}$-optimal design for estimating the parameter $K^T\theta$ in a dose response model with an active control, where the matrix $K$ is given by
$K = \big((\tfrac{\partial}{\partial \theta_1} d^* (\theta))^T, (\tfrac {\partial}{\partial \theta_2} d^*(\theta))^T \big)^T$.
In particular, Theorem \ref{thm2} is applicable and locally AC-optimal designs can be derived from the corresponding optimal designs for model \eqref{exp1}.
The following result provides an alternative representation of the criterion \eqref{critpsi} in the case $s_2=1$. As a consequence the design $\tilde \xi$ required
in Theorem \ref{thm2} is a locally $\tilde c$-optimal design in model \eqref{exp1} for a specific vector $\tilde c$, i.e. the design minimizing $\tilde c^T M_1^-(\tilde \xi, \theta_1) \tilde c$,
where $\tilde c = \tfrac{\partial}{\partial \theta_1} \eta(d^*,\theta_1)$.

\medskip

\begin{Theorem} \label{thm2A}
In the case $s_2=1$, the function in \eqref{critpsi} can be represented as
\begin{equation*} \label{critalt}
\psi(\xi,\theta) = \frac {(\tfrac {\partial}{\partial \theta_2}d^*(\theta))^2}{(\tfrac {\partial}{\partial \theta_2}k(\theta_2))^2} \Bigl \{ \tfrac {1}{1-w_{k+1}} (\tfrac {\partial}{\partial \theta_1}\eta(d^*, \theta_1))^T M^-_1 (\tilde \xi, \theta_1) (\tfrac {\partial}{\partial \theta_1} \eta (d^*,\theta_1)) + (\tfrac {\partial}{\partial \theta_2}k(\theta_2))^2 \tfrac {I_2^-(\theta_2)}{w_{k+1}}  \Bigr \}.
\end{equation*}
\end{Theorem}

\medskip

In the following discussion we determine locally AC-optimal designs for several  nonlinear
regression models accounting for an active control by minimizing the criterion
\eqref{critpsi}.

\subsection{Some explicit  results for two-dimensional models}
In this section we  present some examples illustrating  different structures of
locally AC optimal designs. For this purpose we consider the situation where the Fisher information
matrix $I_1(d,\theta_1)$ defined in \eqref{i2} is of the form
\begin{eqnarray}\label{geo1}
I_1 (d,\theta_1) = \left ( \begin{array} {cc}
f(d,\theta_1)f^T(d,\theta_1) & \bm {0} \\
\bm{0} & \Sigma(\theta_1)
 \end{array} \right ) \in \mathbb{R}^{s_1 \times s_1}
\end{eqnarray}
where $f(d,\theta_1)=(f_1(d,\theta_1), f_2(d,\theta_1))^T$ denotes a two-dimensional vector and $\Sigma(\theta_1)$ a $(s_1-2)\times(s_1-2)$ matrix, which does not depend on the dose level. By Theorem \ref{thm2} the locally AC-optimal design can be determined from the design $\tilde \xi^*$ which minimizes
the expression
\begin{equation} \label{copt2}
\tilde c^T M^-_1 (\tilde \xi, \theta_1) \tilde c
\end{equation}
in the class of all designs defined on the dose range $\mathcal{D}$ for the new drug, where the vector is given by $\tilde c = (\tfrac{\partial}{\partial \theta_1} d^*(\theta))^T$. Because
of the block structure of the Fisher information in \eqref{geo1}  (with a lower block not depending on the dose level) we may
 assume without loss of generality that $s_1=2$, that is
\begin{equation} \label{forf}
M_1 (\tilde \xi, \theta_1) = \int_{\mathcal{D}} f(d,\theta_1) f^T (d,\theta_1) d \tilde \xi (d).
\end{equation}
By Elfving's theorem [see \cite{elfving1952}] a design $\tilde \xi^*$ with weights $\tilde w_i^*$ at the points $d^*_i \ (i=1,\dots,k)$ minimizes
the expression
 \eqref{copt2} if and only if there exists a constant $\gamma > 0$ and $\varepsilon_1,\dots,\varepsilon_k \in \{ -1,1 \}$, such that the
  point $\gamma \tilde c$ is a boundary point of the Elfving set
\begin{equation} \label{elfset}
\mathcal{R} = \mbox{conv} \Bigl \{ \varepsilon f (d,\theta_1) \mid d \in \mathcal{D}, \ \varepsilon \in \{ -1,1 \} \Bigr \}
\end{equation}
and  the representation
$\gamma \tilde c = \sum^k_{i=1} \varepsilon_i \tilde w_i^* f (d^*_i,\theta_1)
$
is valid. Note that $\mathcal{R} = conv\{\mathcal{C} \cup (-\mathcal{C})\}$, where the curve $\mathcal{C}$ is defined by $\mathcal{C}=\{f(d,\theta_1) \mid d \in \mathcal{D}\}$.
The structure of the Elfving set $\mathcal{R}$  depends sensitively  on the distributional assumptions and
we now consider several examples  in  the Michaelis-Menten model.

\begin{Example}  ~~~ \\
{\rm
{
Assume that the dependence on the dose
 in model \eqref{exp1} is described by the Michaelis-Menten model, then the vector $f$ in \eqref{forf} has the form $v(d,\theta_1) (\tfrac{d}{\vartheta_2+d},-\tfrac{\vartheta_1 d}{(\vartheta_2+d)^2})^T$, where the function $v$ varies with the distributional assumption.
 \begin{itemize}
\item [(a)]
In the case of normal distributed responses we have  $v(d,\theta_1)=1$ and it follows by an obvious generalization of Theorem \ref{thm2A},
that we have to consider a $\tilde c$-optimal design problem in model \eqref{exp1}, where the vector $\tilde c$ is now given by
$
	\tilde c=\tfrac{\partial}{\partial \vartheta}\eta(d^*,\vartheta) = (\tfrac{d^*}{\vartheta_2 + d^*},-\tfrac{\vartheta_1d^*}{(\vartheta_2 + d^*)^2})^T.
$
	From the left panel of Figure \ref{normandneg} we observe that the line $\{\gamma \tilde c | ~\gamma > 0\} $ intersects the boundary of the Elfving set
  $\mathcal{R}$ at some point $\mathcal{C } \cup (- \mathcal{C}) $, whenever
  $L \leq x^* \leq d^* < R$,
	where
	$$
		x^*=L \lor \tfrac{\sqrt{2}R^2\vartheta_2+(\sqrt{2}-1)R\vartheta_2^2}{2R^2+4R\vartheta_2+\vartheta_2^2}.
	$$
A typical situation is shown for the vector $\tilde c_2$ in the left panel of Figure \ref{normandneg} for $\vartheta_1=\vartheta_2=2, \mathcal{D}= [0.1,50]$. Consequently, Elfvings theorem shows that a one-point design minimizes \eqref{copt2} in this case. An application of Theorem \ref{thm2} yields that the locally AC-optimal design which allocates $\tfrac{\sigma_1}{\sigma_1+\sigma_2} 100\%$ of the patients to dose level $d^*=\eta^{-1}(\Delta,\vartheta)$ for the new drug and the remaining patients to the active control.
On the other hand, if  $L < d^* \leq x^* < R$, the line $\{\gamma \tilde c | ~\gamma > 0\} $ does not intersect the set $\mathcal{C } \cup (- \mathcal{C})$
at the boundary of the Elfving set ${\cal R}$ and the situation is more complicated. A typical situation for this case is shown for the vector $\tilde c_1$ and the locally AC-optimal design allocates $\rho \tilde \omega_1 100\%$, $\rho \tilde \omega_2 100\%$ of the patients to dose levels $x^*$ and $R$ of the new drug, where $\rho=\tfrac{\sqrt{\delta}\sigma_1}{\sqrt{\delta}\sigma_1+\sigma_2}$ and the remaining patients to the active control,
where
\begin{eqnarray} \label{gewform1}
	\tilde \omega_1 &=& \tfrac{v(R,\theta_1) R(R-d^*)(\vartheta_2+x^*)^2}{v(R,\theta_1)R(R-d^*)(\vartheta_2+x^*)^2 + v(x^*,\theta_1)x^*(x^*-d^*)(\vartheta_2+R)^2},
	\end{eqnarray}
	 $\tilde \omega_2= 1-\tilde \omega_1$,
$\delta =\tilde c^T M_1^{-1}(\tilde\xi^*,\theta_1) \tilde c$ and  $ d^* = \eta^{-1}(\Delta,\theta_1)$.

\begin{figure}[h]
\centering
	\subfigure{\raisebox{0.5cm}{\includegraphics[scale=0.6]{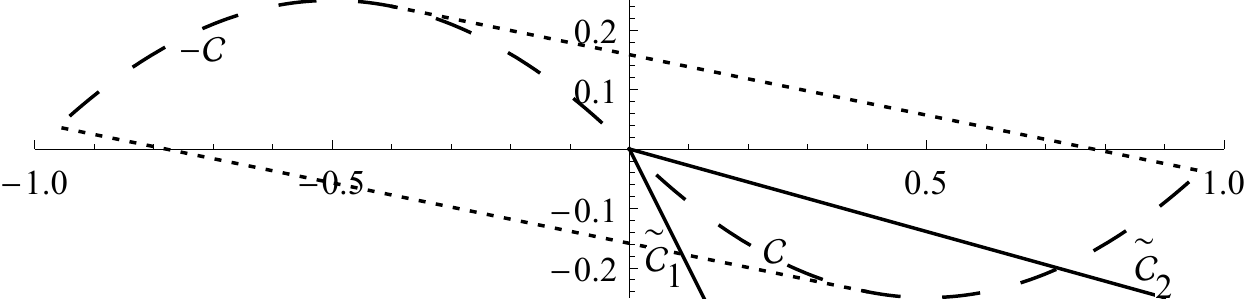}}}
\quad \quad
	\subfigure{\includegraphics[scale=0.5]{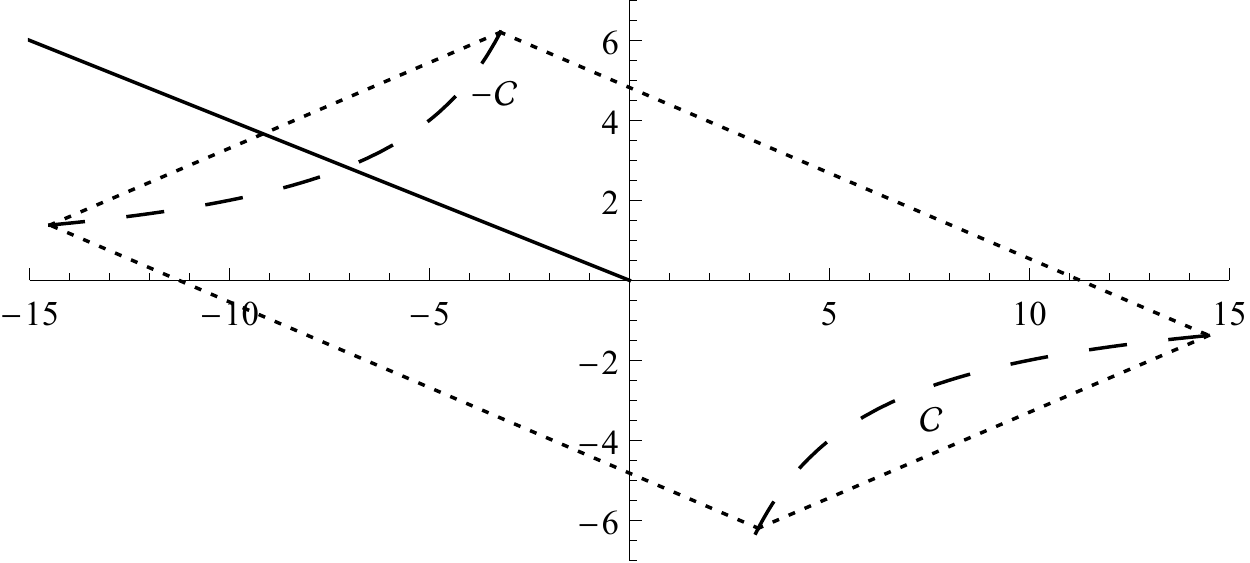}}
\caption{\it The Elfving set  \eqref{elfset}   in model \eqref{exp1}, where the expected response is given by the Michaelis-Menten model. Left panel: normal distribution. Right panel: Negative-binomial distribution.}
\label{normandneg}
\end{figure}

 \item [(b)] As a further example consider the Michaelis Menten model for the probability of a negative binomial distributed response. We have $s_1=2, s_2 =1, $
 $\pi(d,\theta_1)=\tfrac{\vartheta_1 d}{\vartheta_2+d}$, $\tilde c = \tfrac {\partial}{\partial \theta_1} \eta(d^*,\theta_1) = \tfrac{r_1}{\vartheta_1 d^*}(-\tfrac{\vartheta_2+d^*}{\vartheta_1},1)^T$
 and the function $v$ is given by $v(d,\theta_1)=\sqrt{\tfrac{r_1 (d+\vartheta_2)^3}{d^2 \vartheta_1^2 (d(1-\vartheta_1)+\vartheta_2)}}$.
A corresponding Elfving set is depicted in the right panel of Figure \ref{normandneg} for $\vartheta_1=1, \vartheta_2=0.5, \mathcal{D}= [0,10]$ and the locally AC-optimal design is always supported at three points. A straightforward calculation
 shows that the locally AC-optimal design allocates $\rho \tilde \omega_1  100\%$, $\rho \tilde \omega_2 100\%$ of the patients to the dose levels $L$, $R$ for the new drug, where $\rho=\tfrac{\delta \theta_2^2 - \sqrt{(1-\theta_2)\delta \theta_2^2 r_2}}{\delta \theta_2^2- (1-\theta_2)r_2}$ and the remaining patients to the active control, where
\begin{eqnarray}\label{gewform2}
\tilde \omega_1 &=& \tfrac{v(R,\theta_1)R (R-d^*)  (\vartheta_2+L)^2 }{v(R,\theta_1)R (R-d^*) (\vartheta_2+L)^2 + v(L,\theta_1) L (d^*-L)  (\vartheta_2+R)^2  },
\end{eqnarray}
$\tilde \omega_2 = 1-\tilde \omega_1$
$\delta=\tilde c^T M_1^{-1}(\tilde\xi^*,\theta_1) \tilde c$ and $d^* = \eta^{-1}(\Delta,\theta_1)$.

\item [(c)]
Consider now the Michaelis Menten model for binomial distributed responses. We have $s_1=2, s_2 =1, \pi(d,\theta_1)=\tfrac{\vartheta_1 d}{\vartheta_2+d}, \tilde c = \tfrac{\partial}{\partial \theta_1} \pi(d^*,\theta_1)$ and $v(d,\theta_1)=\sqrt{\tfrac{(d+\vartheta_2)^2}{d \vartheta_1 (d (1-\vartheta_1)+\vartheta_2)}}$.
The corresponding Elfving set is depicted in the left panel of Figure \ref{binandpois} for $\vartheta_1=1, \vartheta_2=0.1, \mathcal{D}= [0.02,2]$
 and we have to distinguish three different cases. We observe that the line $\{\gamma \tilde c | ~\gamma > 0\} $ intersects the boundary of the Elfving set
  $\mathcal{R}$ at some point $\mathcal{C } \cup (- \mathcal{C}) $

	if and only if $L \leq x_1^* \leq d^* \leq x_2^* \leq R$,
	where
	$$
		x_1^* = L \lor \tfrac{\vartheta_2(1-\sqrt{1-\pi(R,\theta_1)})}{2\vartheta_1 - 1 + \sqrt{1-\pi(R,\theta_1)}},  \quad x_2^*=
		R \land \tfrac{\vartheta_2(1+\sqrt{1-\pi(R,\theta_1)})}{2\vartheta_1 - 1 - \sqrt{1-\pi(R,\theta_1)}}	.
		$$
A typical situation is shown for the vector $\tilde c_1$ in the left panel of Figure \ref{binandpois}.
Consequently, the same arguments as in the previous examples show that in this case the locally AC-optimal design allocates $\rho 100\%$ of the patients to the dose level $d^*$ of the new drug, where
$\rho=\frac{\delta-\sqrt{ \delta(1-\theta_2)\theta_2}}{\delta- (1-\theta_2) \theta_2}$ and the remaining patients to the active control,
  where $\delta = \tilde c^T M_1^{-1}(\tilde\xi^*,\theta_1) \tilde c$ and $d^*=\eta^{-1}(\Delta,\theta_1)$. \\
 On the other hand, if  $L < d^* \leq x_1^*$, the locally AC-optimal design allocates $\rho \tilde \omega_{11}  100\%$, $\rho  (1-\tilde \omega_{11}) 100\%$ of the patients to dose levels $x_1^*$ and $R$
 of the new drug and the remaining patients to the active control,
where $\tilde \omega_{11}$ is of the form \eqref{gewform1} with $x^*=x_1^*$.
A typical situation is shown for the vector $\tilde c_2$.
The case $L \leq x_2^* \leq d^* \leq R$ corresponds to  the vector $\tilde c_3$. Here
 the locally AC-optimal design allocates $\rho \tilde \omega_{21}  100\%$, $\rho (1-\tilde \omega_{21}) 100\%$ of the patients to dose levels
$x_2^*$ and $R$ of the new drug and the remaining patients to the active control,
where with $L=x_2^*$ $\tilde \omega_{21}$ is of the form \eqref{gewform2} and  $d^*=\eta^{-1}(\Delta,\theta_1)$.

\begin{figure}[h]
\centering
	\subfigure{\includegraphics[scale=0.4]{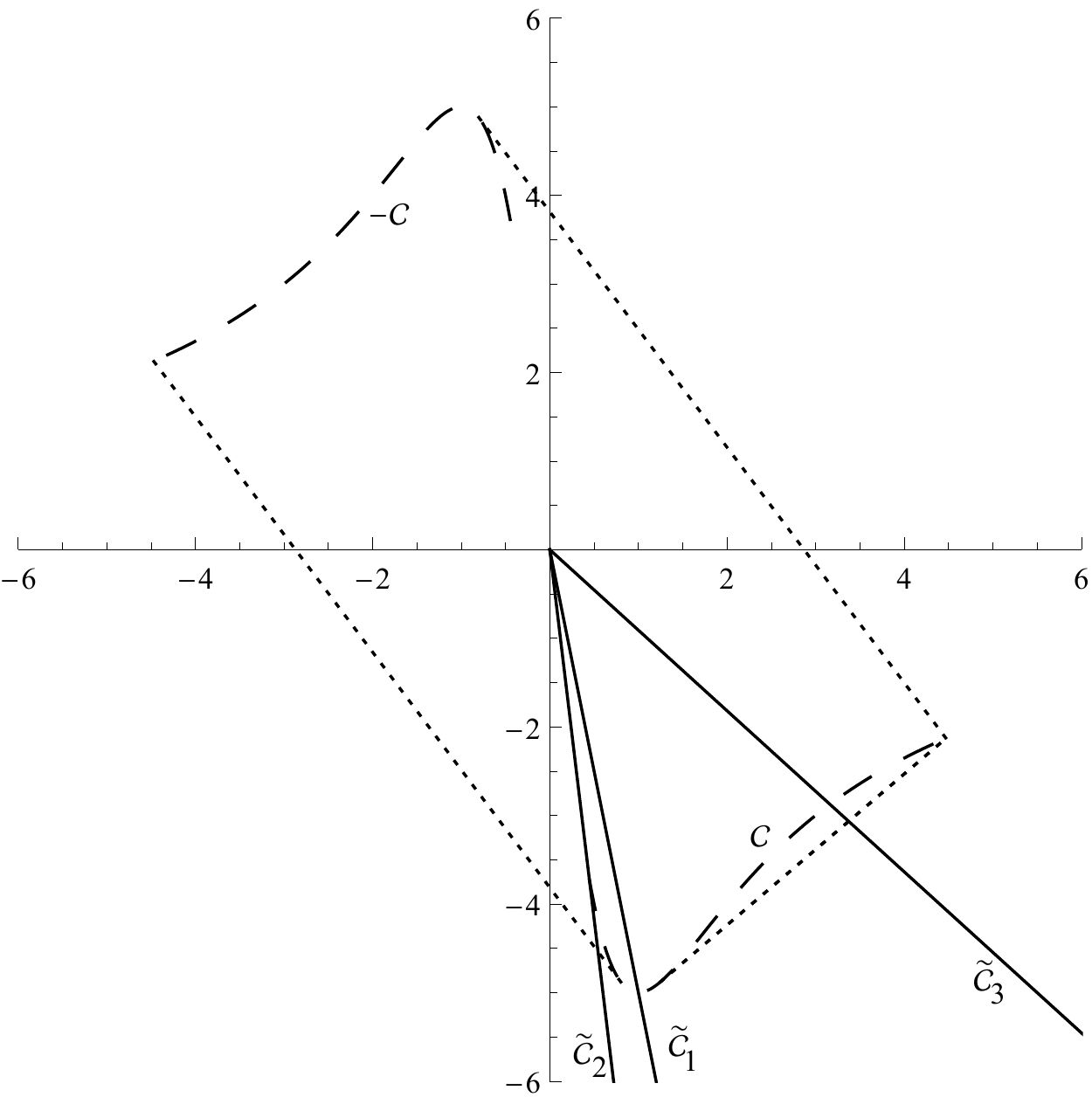}}
	\quad \quad \quad \quad \quad \quad
	\subfigure{\raisebox{0.8cm}{\includegraphics[scale=0.4]{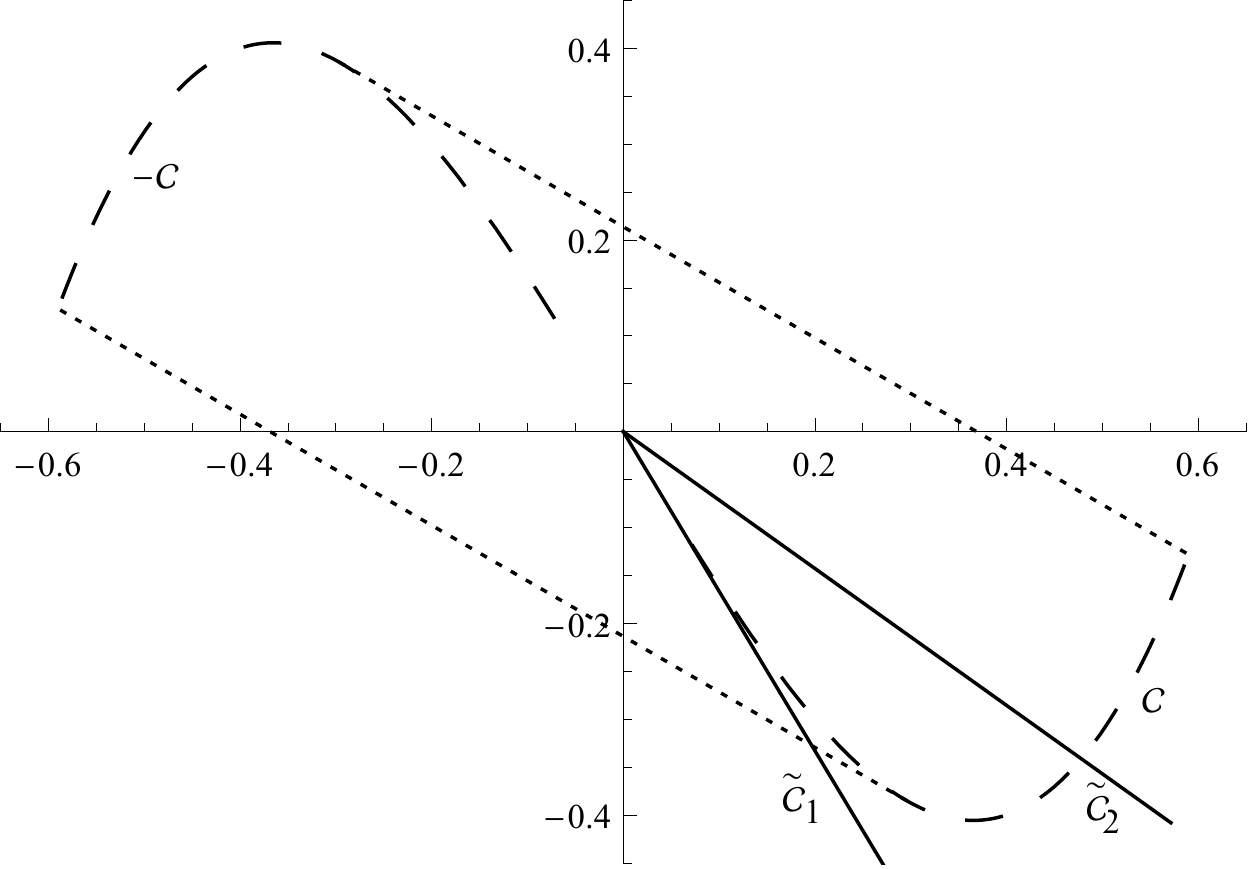}}}
\caption{\it The Elfving set \eqref{elfset} in model \eqref{exp1}, where the expected response is given by the Michaelis-Menten model. Left panel: binomial distribution. Right panel: Poisson distribution.}
\label{binandpois}
\end{figure}

\item [(d)] Finally we consider the case of Poisson distributed responses. We have $s_1=2, s_2 =1, \lambda(d,\theta_1)=\tfrac{\vartheta_1 d}{\vartheta_2+d}$, $v(d,\theta_1)
= {1} \big / {\sqrt{\tfrac{\vartheta_1 d}{\vartheta_2+d}}}$
	and by Theorem \ref{thm2A} we have to solve a $\tilde c$-optimal design problem with $\tilde c = \tfrac {\partial}{\partial \theta_1} \lambda (d^*,\theta_1)= (\tfrac{d^*}{\vartheta_2+d^*},-\tfrac{\vartheta_1 d^{*}}{(\vartheta_2+d^*)^2})^T$. It is easy to see that the line $\{\gamma \tilde c | ~\gamma > 0\} $ intersects the boundary of the Elfving set
  $\mathcal{R}$ at some point $\mathcal{C} \cup (- \mathcal{C}) $ if and only if $L \leq x^* \leq d^* < R$,
	where $x^*=L \lor \tfrac{R \vartheta_2}{3R+4\vartheta_2}$ (see the right panel of Figure \ref{binandpois} for
	$\vartheta_1=2.5, \vartheta_2=1.5, \mathcal{D}= [0.02,10]$ and the vector $\tilde c_2$).
Consequently, the same arguments as in the previous examples show that in this case the locally AC-optimal design allocates $\rho 100\%$ of the patients to dose levels $d^*$ of the new drug, where
$\rho = \tfrac{\sqrt{\delta}}{\sqrt{\delta}+\sqrt{\theta_2}}$ and the remaining patients to the active control,
  where $\delta=\tfrac{d^* \vartheta_1}{\vartheta_2+d^*}$ and $d^*=\lambda^{-1}(\Delta,\theta_1)$.

 On the other hand, if  $L < d^* \leq x^* < R$, the locally AC-optimal design allocates $\rho \tilde \omega_1 100\%$, $\rho (1- \tilde \omega_1)  100\%$ of the patients to dose levels $x^*$ and $R$ of the new drug and the remaining patients to the active control,
where $\tilde \omega_1$ is of the form \eqref{gewform1} with $\delta=(\tfrac{\partial}{\partial \theta_1}\eta(d^*,\theta_1))^T M_1^-(\tilde\xi^*,\theta_1)(\tfrac{\partial}{\partial \theta_1}\eta(d^*,\theta_1))$.
A typical situation is shown for the vector $\tilde c_1$ in the right panel of Figure \ref{binandpois}.
\end{itemize}
}
}

\end{Example}

\subsection{Locally AC-optimal designs in the EMAX model}

Explicit expressions for the AC-optimal designs in the EMAX model are very complicated and for the sake of brevity and better illustration we conclude this paper discussing AC-optimal designs for the two data examples from Section~\ref{sec1}.  \\
We begin with the gouty arthritis clinical trial where we use the same prior information as in Example \ref{dataex}. AC-optimal designs under the  assumption
of a normal and negative binomial distribution  can be found in  the upper part of Table \ref{tabgout}. For example under the assumption of normal distributed endpoints,
the AC-optimal design allocates almost half of the patients to the dose level $101.06$mg and the rest to the active control. In order to compare the standard design introduced in Example \ref{dataex}
we display in the right column the efficiency
\begin{equation} \label{eff}
\eff_{\text{AC}}(\xi,\theta)=\frac{\psi(\xi_{\text{AC}}^{*},\theta)}{\psi (\xi,\theta)} \in [0,1],
\end{equation}
where $\psi(\xi,\theta)$ is defined in \eqref{crit} and $\xi_{\text{AC}}^*$ is the locally AC-optimal design.
\renewcommand{\arraystretch}{1.2}
\begin{table}[h]
\footnotesize
\centering{
\begin{tabular}{|l|c|c|}
\hline
  distribution & AC-optimal design  & $\eff_{AC}$ \\
	\hline
  normal &  \begin{tabular}{cc} $(101.06,0)$  & $(C,1)$ \\ \hline $49.99\%$ & $50.01\%$ \end{tabular} & 0.66 \\
	\hline
	negative binomial & \begin{tabular}{ccc} $(5.44,0)$ & $(300,0)$  & $(C,1)$ \\ \hline $7.6\%$ & $35.6\%$ & $56.8\%$ \end{tabular} & 0.48 \\
	\hline
	\hline
   normal &  \begin{tabular}{cc} $(35.739,0)$  & $(C,1)$ \\ \hline $49.99\%$ & $50.01\%$ \end{tabular} & 0.48 \\
	\hline
	binomial & \begin{tabular}{ccc} $(0,0)$ & $(200,0)$  & $(C,1)$ \\ \hline  $7.34\%$ & $41.95\%$ & $50.71\%$ \end{tabular} & 0.47
 \\
	\hline
 \end{tabular}
 \caption{ \small \it AC-optimal designs  in the two examples from section 1 under different distributional assumptions. Upper part:  gouty arthritis example, with target dose $d^*=100$mg;
 lower part: acute migraine example, with target dose $d^*=35.6$mg. The last column shows the efficiencies of the designs, which were actually used in the study.}
 \label{tabgout}
}
 \end{table}
For example, the efficiency of the standard design for estimating the target dose under the assumption of a normal or negative binomial distribution is $66\%$ and $48\%$, respectively. \\
The second trial is the one in treating migraine and again we use the prior information from Example \ref{dataex}. AC-optimal designs for normal and binomial distributed responses can be found
in the lower part of Table \ref{tabgout}. The efficiencies of the standard design are  given by  $48\%$  and $47\%$ under the assumption of a normal and binomial distribution, respectively.

\bigskip

{\bf Acknowledgements} The authors would like to thank Martina
Stein, who typed parts of this manuscript with considerable
technical expertise.
This work has been supported in part by the
Collaborative Research Center "Statistical modeling of nonlinear
dynamic processes" (SFB 823) of the German Research Foundation
(DFG) and by the National Institute Of General Medical Sciences of the National
Institutes of Health under Award Number R01GM107639. The content is solely the responsibility of the authors and does not necessarily
 represent the official views of the National
Institutes of Health.

\section{Conclusions}

In this paper the optimal design problem for active controlled dose finding studies is considered.
Sufficient conditions are provided such that the optimal design for a dose  finding study with no active control can also be used
for the model with an active control. Our results apply to general optimality criteria and distributional assumptions. In particular they
are applicable in models with discrete responses, which appeared recently in two of our consulting projects. In several examples it is demonstrated that the optimal
designs may depend sensitively on the distributional assumptions. In the clinical trials under consideration these differences were less visible for
$D$-optimal designs. However, in the problem of estimating the target dose  (i.e.
the smallest dose of the new compound which achieves the same treatment effect as the active control), the differences are more substantial, and an optimal design
calculated under a ''wrong'' distributional  assumption (i.e. a normal distribution) might be inefficient, if it used in a different model (i.e. a Binomial model).

\setstretch{1.25}
\setlength{\bibsep}{1pt}
\begin{small}
\bibliographystyle{apalike}
\bibliography{paper23102014}
\end{small}

\normalsize

\section{Appendix: proofs} \label{appendix}
\def\theequation{5.\arabic{equation}}
\setcounter{equation}{0}

{\bf Proof of Theorem \ref{thm1}.}
Assume that the matrix $K$ is a blockdiagonal matrix of the form \eqref{kdiag}. Observing the representations \eqref{inf}, \eqref{kdiag}
and noting that the expression $K^TGK$ is independent of the choice of the generalized inverse of the matrix $M(\xi,\theta)$, we obtain
\begin{eqnarray*}
K^T M^- (\xi,\theta)K= \left(
\begin{array}{cc}
(1-w_{k+1})^{-1} K^T_{11} M^-_1 (\tilde\xi,\theta_1)K_{11} & 0 \\
0 & w^{-1}_{k+1} K^T_{22} I^-_2 (\theta_2)K_{22}
\end{array}
\right).
\end{eqnarray*}
In the case $p \neq 0, -\infty$ this gives for the criterion $\phi_p$ in \eqref{crit} the representation
\begin{eqnarray} \label{crit0}
\phi_p (\xi) &=& \Bigl( \frac {1}{t} \sum^{t}_{i=1} \lambda^{-p}_i (K^TM^-(\xi,\theta)K) \Bigr)^{\frac {1}{p}} \\
&=& \Bigl( \frac {1}{t} \bigl\{
(1-w_{k+1})^{p} \sum^{t_1}_{i=1} \lambda^{-p}_i (K^T_{11} M^-_1 (\tilde \xi, \theta_1)K_{11} \bigr) + w^{p}_{k+1}\sum^{t_2}_{i=1} \lambda^{-p}_i (K^T_{22}I_2^- (\theta_2)K_{22}) \bigr\} \Bigr)^{\frac {1}{p}} \nonumber
 \\ \nonumber
&=& \Bigl( \frac {(1-w_{k+1})^{p}t_1 }{t} (\tilde \phi_p (\tilde \xi))^p + \frac {w^{p}_{k+1}}{t}  tr( (K^T_{22}I_2^-(\theta_2)K_{22})^{-p} ) \Bigr)^{\frac {1}{p}},
\end{eqnarray}
where $\lambda_1(A),\dots,\lambda_n(A)$ denote the eigenvalues of a matrix $A$, $t=t_1+t_2$ and the function $\tilde \phi_p$ is defined in \eqref{crittilde}.

Now it is easy to see that the function $\phi_p$ is an increasing function of $\tilde \phi_p(\tilde \xi)$. Consequently, the locally $\phi_p$-optimal design problem for the dose response model with an active control can be solved by determining  a design $\tilde \xi^*_p$ which maximizes the criterion \eqref{crittilde} in a first step. If $\phi^* = \tilde \phi_p (\tilde \xi^*_p)= \max_{\tilde \xi} \tilde \phi_p(\tilde \xi)$ denotes the optimal value for this criterion, it remains to maximize the function $\phi_p$ in \eqref{crit0} with respect to the weight $w_{k+1}$ assigned to the active control, which gives the expression \eqref{weight} and proves the assertion for the case  $p \neq 0, -\infty$. The remaining cases $p=0$ and $p=-\infty$ are   proved similarly and the details are omitted for the sake of brevity. \hfill $\Box$

\medskip

\textbf{Proof of Theorem \ref{thm31a} } The proof of part (a) has been given in Section \ref{sec3}. For the remaining cases we restrict ourselves to the case of the Poisson distribution for which the Fisher information in model \eqref{exp1} is given by
\begin{equation} \label{31}
I_1 (d,\theta_1) = \frac {d}{\vartheta_1(\vartheta_2+d)} \left (  \begin{array} {cc}
1 & -\frac {  \vartheta_1}{\vartheta_2+d} \\
- \frac {\vartheta_1}{\vartheta_2+d} & \frac {\vartheta_1^2}{(\vartheta_2+d)^2}
\end{array}\right )
\end{equation}
[see equation \eqref{finfpoiss1}]. All other cases are treated similary. By Corollary \ref{cor1} the $D$-optimal design can be obtained from the $D$-optimal design $\tilde \xi^*$ in a regression model with Fisher information \eqref{31}. If $M_1(\tilde \xi, \theta_1) = \int_{\mathcal{D}}I_1(d,\theta_1) d \tilde \xi(d)$ denotes an information matrix of a design $\tilde \xi$ in this model, then $\tilde\xi^*$ is $D$-optimal if and only if the inequality
$
\mbox{tr} (I_1(d,\theta_1)M^{-1}_1(\tilde \xi^*,\theta_1)) \leq 2
$
holds for all $d \in \mathcal{D}$ (see Lemma \ref{lem1}). Moreover, there must be equality at the support points of the design $\tilde \xi^*$. It is easy to see that this inequality is equivalent to an inequality of the form $P_3(d) \leq 0$ where $P_3$ is a polynomial of degree $3$ with $P(0)<0$. A straightforward argument now shows that $\tilde \xi^*$ has exactly two support points $d^*_1 > 0$ and $d^*_2 = R$. Consequently, the $D$-optimal
 design $\tilde \xi^*_1$ for the regression model with information matrix \eqref{31} has equal masses at the points
$d^*_1$  and $ R $,
where $d^*_1$ maximizes the function
$$
f(d) = \frac {R(R-d)^2d}{4(R+\vartheta_2)^3 (\vartheta_2+d)^3}
$$
in the interval $[L,R]$, that is $d^*_1= L \vee \frac {\vartheta_2R}{3\vartheta_2 + 2R}$.
The assertion now follows by an application of Corollary \ref{cor1}, observing that $t_1=2, t_2=2$ in the case under consideration.

\medskip

\textbf{Proof of Thorem \ref{thm2A}} Note that $\Delta = k(\theta_2)$ and that the dose level $d^*(\theta) = \eta^{-1}(\Delta, \theta_1)$ can be defined as the (unique) solution of the equation
$
F(d,\theta) = k(\theta_2) - \eta(d,\theta_1) = 0
$
with respect to $d$. Consequently, the implicit function theorem shows that the function $\theta \to d^*(\theta)$ is differentiable with respect to $\theta$ with gradient given by
$$
\big( (\tfrac {\partial}{\partial \theta_1} d^*(\theta))^T,  \tfrac {\partial}{\partial \theta_2} d^*(\theta) \big)^T = - \big( \tfrac{\partial}{\partial d} F(d,\theta) \Big|_{d=d^* (\theta) }\big)^{-1}
\big( - (\tfrac {\partial}{\partial \theta_1} \eta(d,\theta_1))^T, \tfrac {\partial}{\partial \theta_2} k(\theta_2) \big)^T \Big|_{d=d^*(\theta)},
$$
which implies (comparing the second components) $- \big( \tfrac{\partial}{\partial d} F(d,\theta) \Big|_{d=d^* (\theta) }\big)^{-1} = {\tfrac {\partial}{\partial \theta_2} d^*(\theta)}
\big / {\tfrac {\partial}{\partial \theta_2} k(\theta_2)}. $
Altogether this gives for the first component
\begin{eqnarray*}
(\tfrac {\partial}{\partial \theta_1} d^*(\theta))^T &=& \big( \tfrac{\partial}{\partial d} F(d,\theta) \Big|_{d=d^* (\theta) }\big)^{-1}( \tfrac {\partial}{\partial \theta_1} \eta(d^*,\theta_1))^T = - \frac {\tfrac {\partial}{\partial \theta_2}d^*(\theta)}{\tfrac{\partial}{\partial \theta_2}k(\theta_2)} (\tfrac {\partial}{\partial \theta_1} \eta(d^*,\theta_1))^T
\end{eqnarray*}
and the result follows from the representation \eqref{crit}.

\end{document}